\documentclass[12pt]{article}
\usepackage[english]{babel}
\usepackage{amssymb,latexsym,epsfig,cite}
\usepackage{graphicx}
\usepackage{fix-cm}

\newcommand{\be}{\begin{equation}}
\newcommand{\ee}{\end{equation}}
\def\bea{\begin{eqnarray}}
\def\eea{\end{eqnarray}}

\begin{document}


\begin{center}
\Large{\bf Extended WKB method, resonances and supersymmetric
radial barriers}
\end{center}

\vskip2ex
\begin{center}
Nicol\'as Fern\'andez-Garc\'ia${}^1$ and Oscar Rosas-Ortiz${}^2$\\[2ex]
${}^1$ {\em Instituto de F\'isica, UNAM,\\
AP 20-353, 01000 M\'exico D.F., Mexico\\[1ex]
${}^2$ Departamento de F\'{\i}sica, Cinvestav,\\
AP 14-740, 07000 M\'exico~DF, Mexico}
\end{center}

\begin{center}
\begin{minipage}{14cm}
{\bf Abstract.} Semiclassical approximations are implemented in
the calculation of position and width of low energy resonances for
radial barriers. The numerical integrations are delimited by
$\tau/\tau_{\rm life}\ll 8$, with $\tau$ the period of a classical
particle in the barrier trap and $\tau_{\rm life}$ the resonance
lifetime. These energies are used in the construction of `haired'
short range potentials as the supersymmetric partners of a given
radial barrier. The new potentials could be useful in the study of
the transient phenomena which give rise to the Moshinsky's
diffraction in time.
\end{minipage}
\end{center}

\section{Introduction}
\label{intro}

The study of resonant scattering processes has received
considerable attention in contemporary physics. Resonances are
experimentally observed in atomic, nuclear, and particle physics
so that diverse theoretical models have been proposed for their
analysis over the years (see, e.g. \cite{Bra89,Ros08}). In a
simple picture, a resonance is a special case of scattering state
for which the `capture' of the incident wave produces delays in
the scattered wave. The `time of capture' can be connected with
the lifetime of a decaying system composed by the scatterer and
the incident wave. Then, the resonance state is represented by a
solution of the Schr\"odinger equation associated to a complex
eigenvalue $\epsilon$ and satisfying purely outgoing conditions
(Siegert functions) \cite{Sie39}. These functions are not finite
at $r\rightarrow \infty$, so that they are not admissible as
physical solutions into the mathematical structure of quantum
mechanics. Some approaches extend the formalism of quantum theory
to the wider background where the Siegert functions are more than
a convenient model to solve scattering equations
\cite{Boh89,Del02,Civ04}. Notwithstanding, the `unphysical'
behavior of the Siegert functions has been relevant in the
construction of complex supersymmetric partners of a given
potential \cite{Ros07,Fer08,Cab08} (see also \cite{Fer03,Ros03}).
The main problem is to evaluate the complex point $\epsilon$ up to
a reasonable precision; the binding energy $E=\mbox{Re}(\epsilon)$
and the lifetime $\tau_{\rm life}=-1/(2\mbox{Im}(\epsilon))$ of
the decaying composite are then automatically determined. Below,
the derivation of analytical expressions for $E$ and $\tau_{\rm
life}$ is discussed such that integrations are achievable up to
the precision delimited by $\tau/\tau_{\rm life}\ll 8$, with
$\tau$ the period of a classical particle in a barrier trap. The
results are used to get Darboux-deformations of radial barriers
presenting `hair' over the top. A stronger resonant phenomenon is
expected to be associated with these new interactions.

The avoiding of the asymptotic divergence of the Siegert functions
is usually faced with $r$-complex coordinates $r=\rho
e^{i\lambda}, \lambda >0$. The Siegert functions are then easily
obtained by numerical integration, though they become
eigenfunctions associated to complex eigenvalues of a
non-Hermitian Hamiltonian, as discussed in Sect.~\ref{sec:2}. To
improve the numerical approximations it is feasible to include
variations in the complex $r$-plane \cite{BBJS}, as is briefly
outlined in Sect.~\ref{sec:2a}. The latter approach, however,
depends on one's ability to guess a trial function as a reasonable
approximation to the actual wave function \cite{Tay06}. Variants
of the WKB method are also available to get a more direct
mechanism of integration \cite{Mil53,Kua93,Mur93}. Indeed, in most
cases, bound and resonance energies correspond to poles of the
$S$-matrix, since they are zeros of the Jost function involved.
Such connection suggests that the bound energy techniques can be
extended to the resonance case. Section~\ref{sec:3} deals with the
application of semiclassical approximations to calculate $E$ and
$\Gamma$ in the one-channel, $s$-wave situation. The embedding of
the energies into the complex $\epsilon$-plane due to small
variations of the related wave-function is necessary to ensure the
analytical continuation to the lower half-plane. As a test, the
approach is applied to the simple scattering problem reported in
\cite{BBJS}. Our results are in good agreement with those reported
in e.g. \cite{BBJS,Kor82}. Applications to more elaborated
potentials (as those reported in \cite{Mur93}) are
straightforward. The implementation of simple and double complex
Darboux transformations is discussed in Section~\ref{sec:4}.
Interestingly, `haired' barriers are found to be the
supersymmetric partners of the conventional radial ones.
Connections to the time delay problem in quantum mechanics are
transparent, so that the haired barriers could be of interest in
the studying of the Moshinsky's diffraction in time. The paper is
closed with some concluding remarks.

\section{Resonances and radial barriers}
\label{sec:2}

Consider a spinless particle in a spherical, suitably short range
potential $U(r)$. In dimensionless form, the reduced radial
Schr\"odinger equation is given by
\be
\left[ \frac{d^2}{dr^2} -\frac{\ell(\ell +1)}{r^2} -V(r) +k^2
\right] \psi(r)=0
\label{schro1}
\ee
where $R(r)=\psi(r)/r$ is the radial part of the complete solution
$\phi(\vec r)$. When $r\rightarrow \infty$, the effective
potential vanishes and $\psi$ behaves like combinations of $e^{\pm
i kr}$. Let us take $k =\kappa e^{i\alpha} \in \mathbb C$, $\alpha
\in \mathbb R$, $\kappa:= \vert k \vert$ and $r\geq 0$ to write
$ikr= i\kappa r\cos{\alpha}-\kappa r\sin{\alpha}$. Henceforth the
plane wave functions
\[
\varphi_{\pm}(r,\kappa,\alpha)= e^{\pm ikr}= e^{\pm i\kappa
r\cos{\alpha}}e^{\mp \kappa r\sin{\alpha}}
\]
can be analyzed in terms of $\alpha$. Three general cases are
distinguishable: {\em Scattering energies}. If $\alpha =0$ then $k
=\kappa$ and the complex functions $\varphi_{\pm} = e^{\pm i
\kappa r}$ oscillate with finite amplitude at large distances.
{\em Bound energies}. If $\alpha=\pi/2$ one gets $k=i\kappa$, and
the real function $\varphi_-$ is divergent, while $\varphi_+
=e^{-\kappa r}$ becomes zero for large values of $r$. {\em
Resonance energies}. If $\alpha = -\beta$, with $0<\beta
<\frac{\pi}{2}$, then $k$ is in the fourth quadrant of the complex
$k$-plane and the oscillation amplitude of the complex function
$\varphi_-$ ($\varphi_+$) decreases (increases) exponentially as
$r\rightarrow +\infty$. In any case, the solutions of
(\ref{schro1}) that actually behave like $\varphi_{\pm}$ are the
Hankel functions $h_{\ell}^{\pm}(z)$, for which $h_{\ell}^{\pm}(z)
\rightarrow e^{\pm i(z-\ell \pi/2)}$ as $z\rightarrow \infty$
(see, e.g., \cite{Tay06}). Therefore, solutions corresponding to
bound or to resonance states are picked out from the linear
combinations of $h_{\ell}^{\pm}(z)$ such that only outgoing waves
exist. That is, these solutions satisfy the Siegert condition,
\be
\lim_{r\rightarrow +\infty} \frac{1}{\psi(r)} \frac{d}{dr} \psi(r)
= ik, \qquad k\in \mathbb C.
\label{pureout}
\ee
Thereby both bounded and resonance functions behave as $\varphi_+$
at large distances. Clearly, these solutions must also be regular
at the origin and smooth in between.

\subsection{Complex-scaling method revisited}
\label{sec:2-1}

Let us take the transformation $H \rightarrow UHU^{-1} =
H_{\theta}$, with $\theta$ a dimensionless parameter and the
operator $U$ such that $r\rightarrow re^{i\theta}$. Then $ikr
\rightarrow ikre^{i\theta}$ and the bound wave-functions
$\varphi_+ (r,i\kappa, \frac{\pi}{2})$ become exponential
decreasing complex functions for large values of $r$ provided that
$-\frac{\pi}{2}< \theta < \frac{\pi}{2}$. In turn, the Siegert
functions are transformed as follows
\be
\varphi_+(r,k,-\beta) \rightarrow e^{i\kappa r \cos (\theta -
\beta)} e^{-\kappa r \sin (\theta -\beta)}, \qquad
0<\beta<\frac{\pi}{2}.
\label{siegert}
\ee
Thereby, $\varphi_+(r,k,-\beta)$ is mapped into a bounded function
if $\theta-\beta>0$, i.e. for $\theta \in (0,\frac{\pi}{2})$. As
regards the scattering states, we have:
\[
e^{\pm i\kappa r} \rightarrow e^{\pm i \kappa r\cos(\theta)}
e^{\mp \kappa r \sin(\theta)}.
\]
To preserve the original form of scattering wave-functions the
kinetic parameter $k =\kappa$ must be also modified, thus $\kappa
\rightarrow \kappa e^{-i\theta}$. This last transformation induces
a rotation of the positive real axis of energies in the clockwise
direction by the angle $2\theta: E\propto k^2 \rightarrow \kappa^2
e^{-i2\theta} \propto Ee^{-i2\theta}$. That is, the rotated
scattering energy is complex, $\epsilon = \epsilon_R -i
\frac{\Gamma}{2}$, with $\epsilon_R=E\cos (2\theta)$ and
$\frac{\Gamma}{2}=E \sin (2\theta)$. In summary, the scaling
operator $U$ produces an embedding of the energies $E \in \mathbb
R$ into the complex $\epsilon$-plane by a rotation of $E\geq 0$
such that:

\begin{itemize}
\item[(1)]
The bound state energies are preserved
\item[(2)]
The cut is rotated downward making an angle $2\theta$ with the
real axis, and
\item[(3)]
The resonances are {\em exposed} by the cut.
\end{itemize}

\noindent
In general, the new `complex eigenvalues' are
$\theta$-independent, so that the resonance phenomenon is
associated to the discrete part of the complex-scaled Hamiltonian
$H_{\theta}$. Finally, considering the operators $R$ and $\Pi$,
with $[R,\Pi]= iI$, one realizes that $U=e^{-\theta R\Pi}$ is the
scaling operator we are looking for. Indeed, since $H_{\theta}=
e^{-2i\theta} \Pi^2 + V(e^{i\theta}R)$ is such that
$H_{\theta}^{\dagger} \neq H_{\theta}$, the `regularized' Siegert
functions (\ref{siegert}) are square-integrable eigenfunctions
associated to complex eigenvalues $\epsilon$ of the non-Hermitian
Hamiltonian $H_{\theta}$.

\subsection{BBJS complex-coordinate analysis}
\label{sec:2a}

The repulsive potential
\be
V(r)=V_0 r^2e^{-\lambda r}
\label{suku}
\ee
represents a radial barrier of maximum height $V_{\rm max}=
4V_0/(\lambda e)^2$ at $r_0=2\lambda^{-1}$, which admits narrow
shape resonances in low-energy scattering. Introduced in 1974 by
R. A. Bain, J. N. Bardsley, B.R. Junker and C. V. Sukumar (BBJS),
this potential is widely used in the testing of diverse approaches
developed to calculate resonances (see e.g.
\cite{BBJS,Kor82,Gaz76,Isa78,Mai80}). In their paper, Bain and
coworkers present a combination of the complex-scaling method with
variational principles to analyze the resonance phenomenon.
Indeed, they found a single resonance $\epsilon_0 = 6.8722 -i
0.025549$ (a factor $2$ must be considered in \cite{BBJS}) for the
potential (\ref{suku}) with $\lambda =1$ and $V_0 = 15$. To depict
the BBJS procedure consider equation (\ref{schro1}). After a
complex scaling transformation, we get $M_{\theta} \psi \equiv
e^{2i\theta}[H_{\theta}-k^2]\psi=0$. The next step is to assume
that the integral
\[
I = \int_0^\infty \psi(r) M_{\theta} \psi(r) dr
\]
is preserved for diverse values of $\theta$ and small variations
of $\psi(r)$. Introducing a trial function $f_\epsilon$,
parameterized by a set of variable numbers $c_1, \ldots, c_n$, the
approximation $f_\epsilon(r; c_1, \ldots, c_n)\approx \psi(r)$ is
better for small variations of $I$ with respect to each of the
parameters. That is, every one of the expressions $\frac{\partial
I}{\partial c_i} \approx 0$ becomes an equality if $f_\epsilon =
\psi$. To state precisely: the BBJS method allows for the
determination of approximate values of $k$, leading to points
$\widetilde \epsilon =k^2 \in \mathbb C$ which are in the vicinity
of the resonances $\epsilon = E-i\frac{\Gamma}{2}$ we are
interested in.

\section{The WKB method for resonances}
\label{sec:3}

\begin{figure}[htb]
\centering\includegraphics[width=7cm]{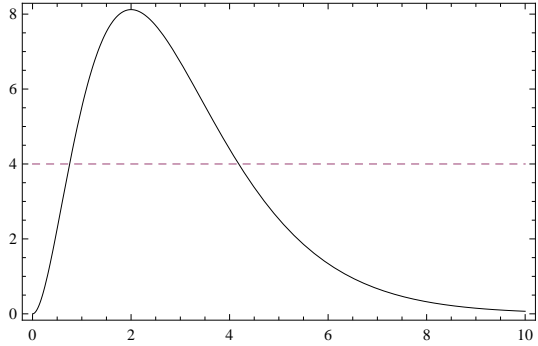}

\caption{\small A typical short-range radial barrier ({\em solid
curve}). For $E$ given ({\em dashed line}), the classical turning
points define two classically accessible regions $(E>V)$, in the
sequel labelled 1 and 3 from {\em left} to {\em right}, and a
nonclassical domain ($E<V$) labelled hereafter region~2.}
\label{fig:1}
\end{figure}

In this section we formulate an extension of the WKB method to
include the calculation of complex-valued energies $\epsilon = E
-i\frac{\Gamma}{2}$. For simplicity, we shall focus on low energy
$s$-wave shape resonances. We use the abbreviation
$p(r)=\sqrt{\vert T \vert}$, with $T=E-V(r)$, the {\em kinetic
parameter} for a given energy $E$. The WKB wave function in a
classically accessible region $(T>0)$ and in the nonclassical
domain $(T<0)$ is, respectively, written as follows
\be
\psi(r) =\frac{1}{\sqrt{p(r)}} e^{\pm i W(c,r)}, \qquad \psi(r)
=\frac{1}{\sqrt{p(r)}} e^{\pm \Omega(c,r)},
\label{wkbfunc}
\ee
where
\be
\displaystyle W(c,r) =\int_c^r\, p(r) \,dr, \qquad \Omega(c,r) =
iW(c,r).
\label{omegas}
\ee
If $p(a)=p(b)=0$, the roots $r=a$ and $r=b$ are the classical
turning points for energies below the barrier (see
Figure~\ref{fig:1}). The origin is a fixed turning point so that
$(0,a)$ and $(b,+\infty)$ are the classically allowed regions
while $(a,b)$ is the nonclassical domain. The connection formulae
can be summarized as follows
\be
\frac{1}{\sqrt{p(r)}} \cos\left( W(c,r)- \frac{\pi}{4} + w \right)
\leftarrow \psi\rightarrow \frac{\sin{w} \, e^{\Omega(r,c)}}
{\sqrt{p(r)}}+ \frac{1}{2}\frac{\cos{w}
\,e^{-\Omega(r,c)}}{\sqrt{p(r)}}
\label{conecta}
\end{equation}
where $w$ is a real parameter defined by the turning point which
is under analysis. We want to get the quantization rule for the
resonance energy $\epsilon = E- i\frac{\Gamma}{2}$ by imposing the
Siegert condition and looking for regular solutions at the origin.

Let the solution in the nonclassical domain be written in the form
\be
\varphi_2(r) = \frac{A}{2\sqrt{p(r)}}e^{-\Omega(a,r)} +
i\frac{B}{2\sqrt{p(r)}}e^{-\Omega(r,b)},
\label{inner2}
\ee
with $A$ and $B$ constants. The additivity of the
$\Omega$-functions $\Omega(a,b) = \Omega(a,r) + \Omega(r,b)$ leads
to
\be
\varphi_{2}(r) = \frac{A}{2\sqrt{p(r)}} e^{-\Omega(a,b)}
e^{\omega(r,b)}+ i\frac{B}{2\sqrt{p(r)}}e^{-\Omega(r,b)}.
\label{inner3}
\ee
This last solution is connected to that of region~3 via
(\ref{conecta}) with $w =\frac{\pi}{2}$:
\be
\varphi_2 \rightarrow \frac{2i B}{\sqrt{p(r)}} \left( \cos
\left[W(b,r) - \frac{\pi}{4} \right] + \frac{iA}{4B}
e^{-\Omega(a,b)} \sin \left[ W(b,r) - \frac{\pi}{4} \right]
\right).
\label{sol1}
\ee
Now, imposing $4B = A e^{-\Omega(a,b)}$ we arrive at the plane
wave
\be
\varphi_3(r) = i \frac{A}{2\sqrt{p(r)}} \exp\left\{-\Omega(a,b) +i
\left[W(b,r)- \frac{\pi}{4} \right] \right\}.
\label{sol2}
\ee
In a similar manner, the connection to region~1 $(w =0)$ leads to
\be
\varphi_1(r)= \frac{A}{\sqrt{p(r)}} \left( \cos \left[ W(r,a) -
\frac{\pi}{4} \right] - \frac{i}{4} e^{-2\Omega(a,b)} \sin \left[
W(r,a) - \frac{\pi}{4} \right] \right).
\label{sol3}
\ee
At the origin one must get $\varphi_1(0)=0$. However, this
condition is not trivially fulfilled by (\ref{sol3}) so that we
first constrain the $W$-function to satisfy
\be
W(0,a) = \left( n+\frac{3}{4} \right) \pi,
\label{tetabase}
\ee
and then we impose the condition
\be
\left\vert \frac{e^{-2\Omega(a,b)}}{4} \right\vert \ll 1.
\label{cond2}
\ee
That is, we look for a function (\ref{sol3}) such that for small
values of $r$ the real part goes to zero faster than its imaginary
counterpart. The imaginary part of $\varphi_1(r)$ is then expected
to produce the embedding of the energy eigenvalues into the
complex $\epsilon$-plane by the addition of a small (negative)
imaginary term to each of the energies:
\[
\mathbb R \ni E \rightarrow \mathbb C \ni \epsilon =E + \delta E,
\qquad \delta E = -i\frac{\Gamma}{2}, \quad \Gamma>0.
\]
This embedding is relevant since $W$ is displaced to $W + \delta
W$, where
\be
\delta W(0,a) = \left[ \frac12 \int_0^a p^{-1}(r) dr \right]
\delta E \equiv \frac{\tau}{4}\delta E = -i \frac{\tau \Gamma}{8}.
\label{deltaw}
\ee
Here $\tau$ corresponds to the period of a classical particle
moving harmonically from the origin to the turning point $a$.
Thus, $\tau$ is a quadrature depending on the constants of
integration $E$ and $a$. Assuming $\delta W(0,a) \ll 1$, the
variation of equation (\ref{sol3}) shows that $\delta
\varphi_1(r)$ is the correction for $\varphi_1(r)$ to be zero at
the origin of the complex plane. In other words,
$\varphi(0)+\delta \varphi(0)=0$ provided that
\be
\displaystyle \delta W(0,a) + i \frac{e^{-2\Omega(a,b)}}{4}=0.
\label{casi}
\ee
From ({\ref{deltaw}) we finally get an analytical expression to
calculate the resonance width:
\be
\frac{\Gamma}{2}= \frac{e^{-2\Omega(a,b)}}{\tau}.
\label{final}
\ee
The time spent by a trapped quantum in the tunnelling of the
radial barrier is therefore given as follows:
\[
\tau_{\rm life} = \frac{1}{\Gamma} = \frac{\tau}{2}
e^{2\Omega(a,b)} \equiv 2 \left( \frac{\partial W(0,a)}{\partial
E} \right)e^{2\Omega(a,b)}.
\]
Remark that our approach is useful for $0< \tau \Gamma \ll 8$;
otherwise (\ref{cond2}) is not satisfied. Notice also that this
result is consistent with our assumption that $\delta W(0,a) \ll
1$. Now, the turning points $a$ and $b$ move farther apart as
$\tau \rightarrow 0$ and, as a consequence, the factor
$e^{-2\Omega(a,b)}$ decreases exponentially. Therefore $\Gamma$ is
narrower as $a \rightarrow 0$. On the other hand, $a$ and $b$ move
closer as $E\rightarrow V_{\rm max}$, so that $\Gamma$ increases
as $a\rightarrow b$ up to $E\approx V_{\rm max}$, where the simple
WKB method does not apply \cite{Mil53}. For energies $E>V_{\rm
max}$, the turning points diverge into complex values and the
approach is not directly applicable (see, however, \cite{Mur93}
and \cite{Kor82}).

In summary, our method gives better results for energies which lay
deep with respect to $V_{\rm max}$; improvements can be achieved
by considering higher-order phase integral methods \cite{Fro65}.
Quite remarkably, the approach presented here could be applied in
the study of time delay where transient effects are known to be
relevant \cite{Mos51}. That is, the method could be useful in
giving a quantum definition to the difference between the time to
traverse the barrier and the time of going from $a$ to $b$ as a
free particle. Particularly if the energy $E$ is below $V_{\rm
max}$, in which case, classically the particle could not arrive at
the region~3 \cite{Gar93} (see also \cite{Gar95}).

\begin{table}

\begin{center}
\begin{tabular}{rccc}
\hline\noalign{\smallskip}
& BBJS & KLM & Extended WKB\\
\noalign{\smallskip}\hline\noalign{\smallskip}

$15[08.1201]$ & $6.8722-i2.5549(-2)$ & $06.994-i2.787(-2)$
& $07.01129-i3.71716(-2)$ \\
$30[16.2402]$ &  & $11.104-i1.321(-4)$ & $11.05705-i1.41354(-4)$\\
$45[24.3604]$ &  & $14.288-i6.840(-7)$ & $14.21889-i7.05499(-7)$\\

\noalign{\smallskip}\hline
\end{tabular}

\caption{\small The single resonance $\epsilon_0$ for potential
(\ref{suku}) with $\lambda=1$ and the indicated values of $V_0
[V_{\rm max}]$. A factor 2 must be considered in the data
originally reported by BBJS in \cite{BBJS} and KLM in
\cite{Kor82}.}

\label{tab:1}
\end{center}
\end{table}

\subsection{Application: BBJS potential scattering}
\label{sec:3a}

We have derived two expressions to calculate resonances. Equation
(\ref{tetabase}) corresponds to the WKB quantization and localizes
the position $E$ of energies fulfilling the Siegert condition.
Equation (\ref{final}), on the other hand, defines the width
$\Gamma$ of the resonance up to the precision established by $0<
\tau \Gamma \ll 8$. We can go a step further in our approach by
noticing that potential (\ref{suku}) admits a series of low-energy
resonances, in correspondence with the strength $V_0$. For getting
`low-energy' resonances, we adopt the criterion of selecting those
complex eigenvalues $\epsilon$ whose real part is smaller than, or
equal to $V_{\rm max}$. Thus, for $a=r_0$ such that $V(a)=V_{\rm
max}$, we want $W$ in (\ref{tetabase}) to have a maximum at
$a=r_0$. This condition is that the integrand of $W(0,r_0)$
involve the longest distance between $E$ and $V(r)$ for each point
in $[0,r_0)$. The solution of this extremum problem enables us to
identify the number $n$ of resonances in terms of $V_0$, as
requested. Indeed, the straightforward calculation shows that
potential (\ref{suku}) admits $n$ resonances, provided $V_0$ is
limited as follows:
\be
V_0(n-1) \leq V_0 < V_0(n), \qquad n \in \mathbb N,
\label{nresona}
\ee
where
\be
V_0(n) = \beta_0 \left(n + \frac{3}{4} \right)^2, \qquad \beta_0 =
\left( \frac{\pi e}{4 \gamma}\right)^2, \quad n=0,1,2,\ldots
\label{vcond}
\ee
and
\be
\gamma = \int_0^1 \sqrt{1-z^2 e^{2(1-z)}}dz.
\label{gama}
\ee
A numerical integration gives $\gamma \approx 0.55098$, so that
$\beta_0 \approx 15.014$ and potential (\ref{suku}) admits a
single resonance if $8.44539 \leq V_0 < 45.9804$. The strength
$V_0 =15$ used in the BBJS paper \cite{BBJS} is clearly in this
category. The same potential is also analyzed by H.J. Korsch, H.
Laurent and R. M\"ohlenkamp (KLM) in their study of the Milne's
differential equation \cite{Kor82}. Table~\ref{tab:1} shows the
good agreement of our results with those reported in \cite{BBJS}
and \cite{Kor82} for a single resonance. Notice that $\epsilon_0$
is such that its real part is close to $V_{\rm max}=8.12012$.
Figure~\ref{fig:2} shows the Siegert function obtained as a
numerical solution of the corresponding Schr\"odinger equation for
the extended WKB value of $\epsilon_0$. The closeness of the
single BBJS resonance to $V_{\rm max}$ is also observed for the
`highest' resonance in each of our studied cases. Namely, the
highest resonance $\epsilon_N$ is such that $\mbox{Re}(\epsilon_n)
< \mbox{Re}(\epsilon_N) \, \forall n<N$ and $\mbox{Re}(\epsilon_N)
\approx V_{\rm max}$. For instance, Table~\ref{tab:2} includes the
values of the unique five resonances belonging to potential
(\ref{suku}) for $\lambda =1$ and $V_0=350$. The highest one is
such that $\mbox{Re}(\epsilon_4) \approx V_{\rm max}=189.46939$.
For the lowest resonance, on the other hand, the bigger the
difference $V_{\rm max} -\mbox{Re}(\epsilon_0)$, the narrower the
width $\Gamma_0$, as was noticed in the previous section.
According to Table~\ref{tab:3}, for example, $V_0=350$ is such
that $V_{\rm max} -\mbox{Re}(\epsilon_0) = 143.02639$ and
$\Gamma_0/2 = 4.2321 \times 10^{-36}$. The strength $V_0=15$ gives
in turn $V_{\rm max}-\mbox{Re}(\epsilon_0)=1.10883$ and
$\Gamma_0/2=0.03717$. Under these conditions the classical motion
of a trapped particle takes place for larger periods $\tau_0$ in
the former than in the second case.

\begin{figure}[htb]
\centering\includegraphics[width=7cm]{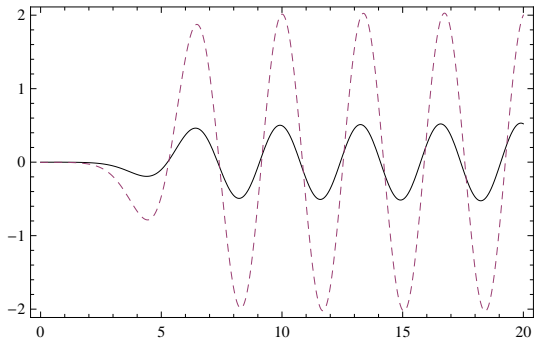}

\caption{\small The real ({\em solid curve}) and imaginary ({\em
dotted curve}) parts of the numerically integrated Siegert
function belonging to the single resonance of $V(r)=15r^2e^{-r}$.}
\label{fig:2}
\end{figure}

As a final illustrative example of the method we include the
anharmonic spherical oscillator $V(r)= \frac12 kr^2 -gr^N$, $k>0,
N=3,5,\ldots$ discussed by Mur and Popov in \cite{Mur93}. This
potential has a barrier of height $V_{\rm max}= \frac{(N-2)}{2N}
k^{N/(N-2)} (gN)^{-2/(N-2)}$ at $r_0^{N-2}=\frac{k}{gN}$.
Considering $s$-waves and the parameters $N=3$, $k=800=2g$
($V_{\rm max}=59.25925$), the potential possesses a single
low-energy resonance at $\epsilon= 47.0105-i0.861937$, which could
be also tested by using the BBJS or the KLM methods. Hotter
resonances for which $E>V_{\rm max}$ can be analyzed with either
the Mur-Popov approach or the KLM one.

\begin{center}
\begin{table}[t]

\begin{center}
\begin{tabular}{cll}
\hline\noalign{\smallskip}
& $\mbox{Re}(\epsilon)$ & $-\mbox{Im}(\epsilon)=\Gamma/2$ \\
\noalign{\smallskip}\hline\noalign{\smallskip}
$\epsilon_0$ & $46.4430$ & $4.23210 \times 10^{-36}$ \\
$\epsilon_1$ & $95.9768$ & $2.91608 \times 10^{-21}$ \\
$\epsilon_2$ & $135.8367$ & $1.23305 \times 10^{-11}$ \\
$\epsilon_3$ & $167.0870$ & $5.98751 \times 10^{-5}$ \\
$\epsilon_4$ & $188.5395$ & $0.71602$ \\
\noalign{\smallskip}\hline
\end{tabular}
\end{center}
\caption{\small Extended WKB values of the unique five resonances
for potential (\ref{suku}) with $\lambda=1$ and $V_0=350$ ($V_{\rm
max} = 189.46939$).}
\label{tab:2}
\end{table}
\end{center}

\begin{center}
\begin{table}[b]
\begin{center}
\begin{tabular}{rcrll}
\hline\noalign{\smallskip} $V_0$ & $n$ & $V_{\rm max}$ &
$\mbox{Re}(\epsilon_0)$ & $-\mbox{Im}(\epsilon_0) =\Gamma_0/2$\\
\hline\noalign{\smallskip}
15 & 1 & $8.12012$ & $7.01129$ & $0.03717$\\
60 & 2 & $32.48046$ & $16.8584$ & $4.86902 \times 10^{-9}$\\
150 & 3 & $81.20116$ & $28.8664$ & $2.14549 \times 10^{-19}$\\
250 & 4 & $135.33528$ & $38.5165$ & $1.97227 \times 10^{-28}$\\
350 & 5 & $189.46939$ & $46.4430$ & $4.23210 \times 10^{-36}$\\

\noalign{\smallskip}\hline
\end{tabular}
\end{center}
\caption{\small The extended WKB values of the `lowest' resonance
$\epsilon_0$ for different strengths $V_0$ of potential
(\ref{suku}) and $\lambda =1$. The number of resonances $n$ is in
correspondence with the condition (\ref{nresona}).}
\label{tab:3}

\end{table}
\end{center}

\section{Susy transformations and concluding remarks}
\label{sec:4}

The Darboux transformation
\be
\widetilde V(r) = V(r) + 2\beta'(r)
\label{darboux}
\ee
is useful in many branches of the mathematical physics (see e.g.
\cite{Rog02}). Of particular interest in quantum theories, this
transformation supports the mathematical structure of the
supersymmetric approach (for recent reviews, see
\cite{Mie04,Bay04,And04,Fer05,Fer09}). Moreover, the axioms for
Hermitian operators and real spectra can be abolished in some
situations (cf.
\cite{Ros07,Fer08,Cab08,Fer03,Ros03,Bay96,And99,Sam06,Mun05,Fer06}
and \cite{Ram03,Sok06}). Two potentials $V$ and $\widetilde V$ are
said to be supersymmetric (Susy) partners if the $\beta$-function
in (\ref{darboux}) is a nontrivial solution of the Riccati
equation,
\be
-\beta'(r) + \beta^2(r) = V(r) -\epsilon.
\label{riccati}
\ee
In such a case, the spectrum of $\widetilde V$ is the same as that
of $V$, occasionally extended by an additional point $\epsilon$
which could be complex. The non-linear equation (\ref{riccati}) is
linearized to the Schr\"odinger equation $H\psi_{\epsilon} =
\epsilon \psi_{\epsilon}$ by means of the logarithmic
transformation $\beta(r) =- \frac{d}{dr} \ln \psi_{\epsilon}(r)$,
with $\psi_{\epsilon}$ not necessarily in ${\cal L}^2(
\mbox{Dom}H$). For $\psi_{\epsilon}$ satisfying the Siegert
condition (\ref{pureout}), one easily verifies that potential
$\widetilde V$ inherits the behavior of $V$ at infinity. It is a
custom to take $\psi_{\epsilon}$-functions with at most a single
zero, provided this root is one of the borders of $\mbox{Dom}(H)$.
In the present case, $\psi_{\epsilon}$ is the Siegert function
belonging to one of the previously derived resonance energies.
Thereby, $\widetilde V$ is complex-valued and singular at the
origin. A second Darboux transformation, using this time
$\widetilde V$ and the complex conjugate of $\beta$, gives rise to
a new Susy partner of $V$, which is a real function. Also this new
potential inherits the initial spectrum and is a regular function
in $\mbox{Dom}H$, as shown in Figure~\ref{fig:3}.

\begin{figure}[htb]
\centering\includegraphics[width=6cm]{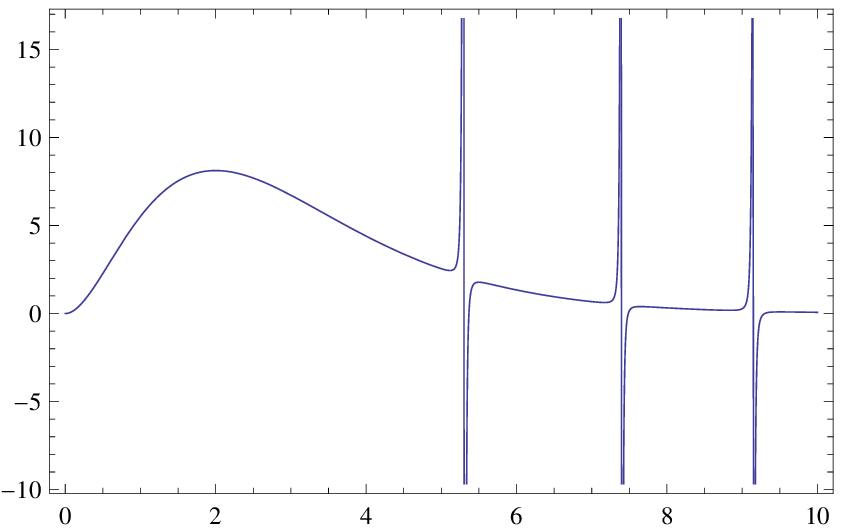} \hskip1cm
\includegraphics[width=6cm]{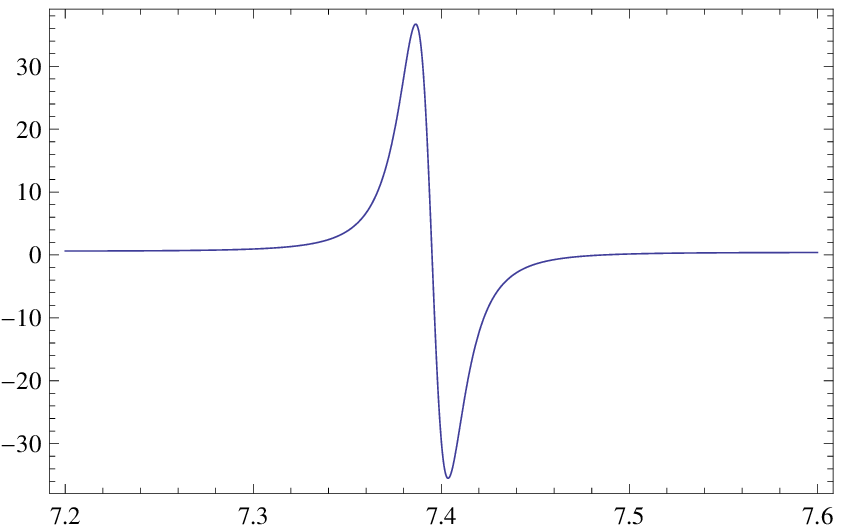}
\caption{\small Haired second order supersymmetric partner of
potential $V(r)=15r^2e^{-r}$. Each of the twice
Darboux-distortions (couple of hairs) is a {\em smooth curve}, as
shown at the {\em right}.}
\label{fig:3}
\end{figure}

The new radial potentials exhibit `hair' along the negative slope,
which induces stronger resonant phenomena. Since the Siegert
function oscillates, while its amplitude increases exponentially
(see Fig.~\ref{fig:1}), for large distances the amplitude of the
$\beta$-function decreases up to $-ik$. Hence, according to the
Siegert condition (\ref{pureout}), the complex double
Darboux-distortions (couple of hairs) are cancelled as
$r\rightarrow \infty$. The same phenomenon is presented in square
wells, where analytical expressions for $E$ and $\Gamma$ have been
obtained and the number of hairs depends on the excitation of the
resonance \cite{Fer08,Fer06}. It is reasonable to assume that
these haired potentials induce delays on the scattering states
which are longer than the delay associated with their
supersymmetric partners. It is then interesting to analyze the
transient phenomena in the scattering process of these new
potentials. In this way, the supersymmetric quantum mechanics
could be connected to the Moshinsky's diffraction in time through
these haired potentials.

\vskip2ex
{\bf Acknownledgementes} This paper is dedicated to Prof. Manolo
Gadella on the occasion of his 60th birthday. ORO is indebted to
Gast\'on Garc\'{\i}a-Calder\'on and Osvaldo Civitarese for
enlightening discussions in {\em El Colegio Nacional}, Mexico. The
support of CONACyT project 24333-50766-F is acknowledged.



\begin{thebibliography}{99}

\bibitem{Bra89}
Br\"andas E and Elander N (Eds), {\em Resonances}; in Lecture
Notes in Physics {\bf 325}, Springer-Verlag, Berlin (1989)

\bibitem{Ros08}
Rosas-Ortiz O, Fern\'andez-Garc\'{\i}a N and Cruz~y~Cruz S, {\em A
Primer on Resonances in Quantum Mechanics}, AIP Conf. Proc. {\bf
1077} 31 (2008)

\bibitem{Sie39}
Siegert AJF, {\em Phys Rev} {\bf 56} 750 (1939)

\bibitem{Boh89}
Bohm A and Gadella M, {\em Dirac kets, Gamow vectors and Gelfand
Triplets}, Lecture Notes in Physics {\bf 348}, Springer-Verlag,
N.Y. (1989)

\bibitem{Del02}
de~la~Madrid R and Gadella M, {\em Am J Phys} {\bf 70} 626 (2002)

\bibitem{Civ04}
Civitarese O and Gadella M, {\em Phys Rep} {\bf 396} 41 (2004)

\bibitem{Ros07}
Rosas-Ortiz O, {\em Rev Mex F\'{\i}s} {\bf 53} (Suppl 2) 103 (2007);\\
Fern\'andez-Garc\'{\i}a N, {\em Rev Mex F\'{\i}s} {\bf 53} (Suppl
4) 42 (2007)

\bibitem{Fer08}
Fern\'andez-Garc\'{\i}a N and Rosas-Ortiz O, {\em Ann Phys} {\bf
323} 1397 (2008)

\bibitem{Cab08}
Cabrera-Mungu\'{\i}a I and Rosas-Ortiz O, {\em J Phys: Conf
Series} {\bf 128} 012042 (2008);\\
Fern\'andez-Garc\'{\i}a N and Rosas-Ortiz O, {\em J Phys: Conf
Series} {\bf 128} 012044 (2008)

\bibitem{Fer03}
Fern\'andez DJ, Mu\~noz R and Ramos A, {\em Phys Lett A} {\bf 308}
11 (2003)

\bibitem{Ros03}
Rosas-Ortiz O and Mu\~noz R, {\em J Phys A: Math Gen} {\bf 36}
8497 (2003)

\bibitem{BBJS}
Bain RA, Bardsley JN, Junker BR and Sukumar CV, {\em J Phys B:
Atom Molec Phys}, {\bf 7} 2189 (1974)

\bibitem{Tay06}
Taylor JR, {\em Scattering Theory. The quantum theory of
nonrelativistic collisions}, Dover, N.Y. (2006)

\bibitem{Mil53}
Miller S and Good R, {\em Phys Rev} {\bf 91}, 174 (1953)

\bibitem{Kua93}
Kuang Y and Gu JS, {\em J Phys B: At Mol Opt Phys} {\bf 26} 3057
(1993)

\bibitem{Mur93}
Mur VD and Popov VS, {\em JETP} {\bf 77}, 18 (1993)

\bibitem{Kor82}
Korsch HJ, Laurent H and M\"ohlenkamo R, {\em J Phys B: At Mol
Phys} {\bf 15}, 1 (1982)

\bibitem{Gaz76}
Gazdy B, {\em J Phys A: Math Gen} {\bf 9} L39 (1976)

\bibitem{Isa78}
Isaacson AD, McCurdy M and Miller WH, {\em Chem Phys} {\bf 34} 311
(1978)

\bibitem{Mai80}
Maier CH, Dederbaum LS and Domcke W, {\em J Phys B: At Mol Phys}
{\bf 13} L119 (1980)

\bibitem{Fro65}
Fr\"omand N and Fr\"oman PO, {\em JWKB Approximation,
contributions to the theory}, North-Holland, Amsterdan (1965)

\bibitem{Mos51}
Moshinsky M, {\em Phys Rev} {\bf 84} 525 (1951); ibid {\bf 88} 625
(1952)

\bibitem{Gar93}
Garc\'{\i}a-Calder\'on G, Mateos JL and Moshinsky M, {\em Rev
M\'ex F\'{\i}s} {\bf 39} (Suppl 2) 76 (1993)

\bibitem{Gar95}
Garc\'{\i}a-Calder\'on G, Mateos JL and Moshinsky M, {\em Phys Rev
Lett} {\bf 74} 337 (1995)


\bibitem{Rog02}
Rogers C and Schief WK, {\em B\"acklund and Darboux
Transformations. Geometry and modern applications in soliton
theory}, CUP, United Kingdom (2002)

\bibitem{Mie04}
Mielnik B and Rosas-Ortiz O, {\em J Phys A: Math Gen} {\bf 37}
10007 (2004)

\bibitem{Bay04}
Baye D and Sparenberg JM, {\em J Phys A: Math Gen} {\bf 37} 10223
(2004)

\bibitem{And04}
Andrianov AA and Cannata F, {\em J Phys A: Math Gen} {\bf 37}
10297 (2004)

\bibitem{Fer05}
Fern\'andez DJ and Fern\'andez-Garc\'{\i}a N, {\em Higher-order
supersymmetric quantum mechanics}, AIP Conf Proc {\bf 744}, 236
(2005)

\bibitem{Fer09}
Fern\'andez DJ, {\em Supersymmetric Quantum Mechanics},
arXiv:0910.0192v1[quant-ph]

\bibitem{Bay96}
Baye D, L\'evai G and Sparenberg JM, {\em Nucl Phys A} {\bf 599}
435 (1996)

\bibitem{And99}
Andrianov AA, Ioffe MV, Cannata F and Dedonder JP, {\em Int J Mod
Phys A} {\bf 14} 2675 (1999)

\bibitem{Sam06}
Samsonov BF and Pupasov AM, {\em Phys Lett A} {\bf 356}  210
(2006)

\bibitem{Mun05}
Mu\~noz R, {\em Phys Lett A} {\bf 345} 287 (2005)

\bibitem{Fer06}
Fern\'andez-Garc\'{\i}a N, {\em On a class of hairy square
barriers and Gamow vectors}, AIP Conf Proc {\bf 885} 30 (2006)

\bibitem{Ram03}
Ram\'{\i}rez A and Mielnik B, {\em Rev Mex F\'{\i}s} {\bf 49}
(Suppl 2) 130 (2003)

\bibitem{Sok06}
Sokolov AV, Andrianov AA and Cannata F, {\em J Phys A: Math Gen}
{\bf 9} 10207 (2006)
\end{thebibliography}
\end{document}